\documentclass[11pt,twoside]{article}
\usepackage{asp2010}
\usepackage{natbib}

\aspvolume{471} 
\aspvoltitle{Origins of the Expanding Universe: 1912-1932}
\aspcpryear{2013} 
\aspvolauthor{Michael J. Way and Deidre Hunter, eds.}

\begin{document}

\title{Slipher's Redshifts as Support for de Sitter's Model and the Discovery of the Dynamic Universe}

\author{Harry Nussbaumer\affil{Institute of Astronomy,
ETH Z\"{u}rich, 8093 Z\"{u}rich, Switzerland}}

\begin{abstract}
Of the first two relativistic world models, only the one by de Sitter
predicted redshifted spectra from far away astronomical objects. Slipher's
redshifts therefore seemed to arbitrate against Einstein's model which made no
such predictions. Both models were trying to describe a static universe.
However, Lema\^{i}tre found that de Sitter's construct resulted in a spatially
inhomogeneous universe. He then opted for a model that corresponded to
Einstein's closed, curved universe but allowed the radius of curvature to change
with time. Slipher's redshifts suggested to him that the universe is dynamic and
expanding. We also discuss the respective merits of Friedman and Lema\^{i}tre in
revealing the dynamic nature of the universe.   
\end{abstract}

\section{Introduction}

Although our conference ``Origins of the Expanding Universe: 1912-1932''
is focused on the contribution of Vesto Slipher, I shall also
discuss the contributions of two theoreticians associated with the discovery of
the expanding universe: George Lema\^{i}tre, the discoverer of the expanding
universe and Alexander Friedman, who was the first to present mathematical
solutions for a dynamical universe.

When Einstein and de Sitter published their cosmological models in
1917, the bulk of their discussion was purely theoretical. It was only as a kind
of afterthought that de Sitter mentioned a possible link between theory and
astronomical observations: Slipher's wavelength shifts observed in spiral
nebulae. I shall therefore first show how Slipher's redshifts seemed to
arbitrate in favor of de Sitter's apparent static and empty universe. Later
they convinced Lema\^{i}tre that we live in an expanding universe. Einstein as
well as de Sitter looked for a static universe that would remain the same forever.
In 1922, Friedman demonstrated the mathematical possibility of a dynamic universe
within the concept of general relativity; however his findings attracted no
attention. In 1927, Lema\^{i}tre, who was not aware of Friedman's work, also found
dynamic solutions. Combining his model with Slipher's observations, he suggested
that we live in an expanding universe.

\section{The evolution in Slipher's attitude towards the line shifts }

Slipher's crucial nebular observations began at a time when the nature
of the nebulae was still an unsolved question. His opinion evolved from viewing
these fuzzy objects as related to planetary nebulae, to being struck by their
large wavelength shifts, and then, after further observation and reflection, he
took his stand on the side of the ``island universe'' faction. The evolution in
his thinking can be seen in his publications from 1912 to 1917
\citep{Slipher1912LowOB...2...26S,Slipher1913LowOB...2...56S,Slipher1915PA.....23...21S,Slipher1917PAPhS..56..403S}.

In 1912, Slipher published a note ``On the spectrum of the nebula in
the Pleiades'' \citep{Slipher1912LowOB...2...26S}.
After a careful discussion of his December 1912
observations he concluded: ``\emph{... that the Andromeda Nebula and similar
spiral nebulae might consist of a central star enveloped and beclouded by
fragmentary and disintegrated matter which shines by light supplied by the
central sun}.'' However, one year later, when analysing his September 1912
observations, he was struck by the large wavelength shift of Andromeda. He had
changed his optical arrangements and had taken a set of spectra with high
resolution, and discovered ``\emph{... that the nebular lines were perceptibly
displaced with reference to the comparison lines}''
\citep{Slipher1913LowOB...2...56S}. He then
concluded: ``\emph{That the velocity of the first spiral observed should be so
high intimates that the spirals as a class have higher velocities than do the
stars and that it might not be fruitless to observe some of the more promising
spirals for proper motion. Thus the extension of the work to other objects
promises results of fundamental importance, but the faintness of the spectra
makes the work heavy and the accumulation of results slow}.'' In 1915 he
published a progress report where he listed 15 nebulae: 3 of them with ``small''
redshifts, 1 with no indication, 2 with negative velocities, and 9 with
positive velocities \citep{Slipher1915PA.....23...21S}. He then mentioned the
\emph{``Campbell-Kapteyn discovery of the increase in stellar velocity with
`advance' in stellar spectral type.''} That hypothesis claimed that the stars at
birth have no motion, but gradually acquire it in passing through their further
development. Slipher remarked that the great nebular velocities would place them
a long way along the evolution; however, he did not dwell on this
interpretation. Then, in April 1917 he  published spectrograms of 25 nebulae, 4
of them  with negative and 21 with positive
velocities \citep{Slipher1917PAPhS..56..403S}.
Slipher now becomes more outspoken about his idea on the nature of spiral
nebulae: ``\emph{It has for a long time been suggested that the spiral nebulae
are stellar systems seen at great distances. This is the so-called ``island
universe'' theory, which regards our stellar system and the Milky Way as a great
spiral nebula which we see from within. This theory, it seems to me, gains
favor in the present observations''}.

Thus, before the ``Great Debate'' of  Curtis and Shapley in 1920, and before
1922, when \"{O}pik placed Andromeda at a distance of 450,000 pc, and before
1925, when Hubble definitely cut the Gordian knot  by resolving Cepheid variable
stars in NGC 6822, M33 and M31, Slipher was convinced of the island universe
hypothesis, because the spirals, as a class, showed very high wavelength shifts,
most of them redshifts, which distinguished them clearly from all other
astronomical objects.

\section{The beginning of modern cosmology}

\subsection{Einstein and de Sitter}

In 1917, Einstein opened a new chapter in cosmology by publishing his
static model of the universe \citep{Eddington1917SPAW.......142E}.
It was generally assumed among
the theoreticians that the universe did not vary in time, and common sense
demanded from any cosmological model that the universe remain static. To comply
with this condition Einstein added the famous cosmological term, $\Lambda $, to
his fundamental equations (I follow the modern notation of a capital $\Lambda $,
whereas in those years it was written as $\lambda $):

\begin{equation}
G_{ij} -\Lambda g_{ij} =-\kappa (T_{ij} -\frac{1}{2} g_{ij} T); \ \
i,j = 1, 2, 3, 4.
\end{equation}

We can attribute the first three indices to the spatial world, and the
fourth to time. For symmetry reasons the set of equations reduces to 10,
however, only six equations are independent. The solution of these equations is
the metric tensor \emph{g${}_{ij}$ }which describes the geometrical structure
of the universe. Assuming also a homogeneous distribution of matter, Einstein
derived a model of the universe that was static and its spatial part is of
closed curvature. His 3-dimensional spatial world can be projected onto a
circle. This circle maintains its radius, the radius of the universe, for all
past and future times. Thus, if the dimension of time is added to the
projection, the model becomes cylindrical; this is called Einstein's cylindrical
world.

A few months after Einstein's publication, the Dutch astrophysicist
Willem de Sitter also derived a cosmic model from Einstein's field equations 
\citep{deSitter1917MNRAS..78....3D}.
However, he made a further, drastic simplification by assuming a
universe empty of matter. Thus his universe was represented by the equation
\begin{equation}
G_{ij} -\Lambda g_{ij} =0,
\end{equation}
where the energy term on the right hand side has been set to zero.

When describing physical events one is, within certain limits, free in
the choice of the coordinate system. For the line element in his 4-dimensional
space-time de Sitter chose the form

\begin{equation}
ds^{2} =R^{2} \left(-d\chi ^{2} -\sin ^{2} \chi
\left(d\theta ^{2} +\sin ^{2} \theta d\phi ^{2} \right)+\cos ^{2} \chi dt^{2}
\right),
\end{equation}

\noindent where $\chi = r / R$; $r$ is the distance from the
observer, $R$ is the radius of curvature. Or, for the propagation of
light we have $ds=0$, and accordingly for constant $\theta$ and
$\phi$

\begin{equation}
dt=\sec \chi d\chi, where  \sec \chi =1/\cos \chi.
\end{equation}

\noindent Time runs slower when $r$ increases. Since the interval
$dt$ between two points in time increases when $r$ increases, the
frequency decreases and the wavelength increases. However, it was later shown
that the model contained a flaw, as shown below.

Einstein did not offer any astronomical observations to verify his
model. However, de Sitter, at the end of his very theoretical treatise, pointed
to its observational implications: ``\emph{\dots we have
$g_{44} =\cos ^{2} \chi $.  Consequently the frequency of light-vibrations
diminishes with increasing distance from the origin of coordinates.
The lines in the spectra of
very distant stars or nebulae must therefore be systematically displaced towards
the red, giving rise to a spurious positive radial velocity}.'' He further
added: ``\emph{Recently a number of radial velocities of these nebulae have
been determined}.'' He referred to a Report to the Council of the RAS in 1917,
where Eddington refers to Slipher's first determination of the radial velocity
of a spiral nebula and to other investigators who confirmed Slipher's
observations \citep{Eddington1917MNRAS..77..375}. De Sitter then mentioned wavelength
shifts of three nebulae -- M31, NGC 1068, NGC 4594 -- and thought that they might
strengthen his model's claim to validity. From their mean recession velocity of
600 km s$^{-1}$ and an assumed mean distance of 100 kpc
(today's accepted distance to
M31 is $\approx$ 800 kpc) he arrived at a radius of curvature of his universe of
$R=3\times10^{11}$ astronomical units, or 1.5 Mpc. But then he added that
this result, derived from only three nebulae, had practically no value. However,
should further observations confirm that the spiral nebulae had systematically
positive radial velocities, this would be a strong indication that his model was
correct.

De Sitter's paper did not stir up great observational activity, but,
as shown by \cite{Nussbaumer2009discovering}, his empty space initiated much
discussion among the theoreticians, in particular Einstein, de Sitter, Klein,
Lanczos and Weyl. 

\subsection{Eddington's book of 1923}

The publication of the book \emph{The Mathematical Theory of Relativity}
\citep{Eddington1923mtr..book.....E} set the observers in motion,
as can be seen from the example of \cite{Wirtz1924AN....222...21W}.
Chapter 70 carries the title ``Properties of de Sitter's
spherical world'' and contains the cosmologically essential points of de
Sitters's theoretical model. It also features a table with 41 radial velocities
of spiral nebulae, measured by Slipher up to February 1922. Eddington mentions
that Slipher had prepared that table for him, inserting many unpublished
results. He also adds some thoughts about the physical meaning of de Sitter's
empty universe: Is it really empty, or has all the matter simply been swept into
a ring of peripheral matter necessary in order to distend the empty region
within? He offers no answer. He dwells on the slowing down of time in objects of
increasing cosmological distances, such that their spectral lines would appear
displaced towards the red. The formula turns out to be

\begin{equation}
\frac{\Delta \lambda }{\lambda } =\frac{1}{2} \left(\frac{r}{R} \right)^{2},
\end{equation}

\noindent where $r$ is the distance to the object and $R$ is the
radius of curvature of the universe. Already in 1924 Eddington's book had its
second edition, and in 1925 it was translated into German. It became a standard
textbook. In the book, Eddington also raised the possibility of a redshift
contribution from the cosmological constant, $\Lambda $, because it acts as an
accelerating force and pushes test particles away from the observer.

Thus, by the middle of the 1920s, Slipher's wavelength-shifts
signaled either motion in a conventional world, or a change in our concept of
time in the sense of de Sitter's universe: the redshifts had important
philosophical implications. 

\section{The observers preoccupation with de Sitter}

\subsection{Wirtz tries to verify de Sitter}

In 1922, Wirtz drew attention to the availability of radial motions of
29 spiral nebulae that seemed to indicate a general dispersal away from us
\citep{Wirtz1922AN....215..349W}. He thought
that they might hold a key to the structure of the
universe. He gave a list of 29 NGC objects with their radial velocities,
collected from different sources, which he did not identify. For the velocities
he found an approximate linearity in the sense that the closer nebulae approach
us, whereas the more distant ones tend to recede. A global look at the data
suggested to him a general expansion of the system of spiral nebulae, and he
remarks that no such tendency is seen in globular clusters. However, Wirtz did
not refer to any theoretical model at this point.

Following the publication of  Eddington's book in 1923, Wirtz
responded with the article ``De Sitter's cosmology and the radial motion of
spiral nebulae'' \citep{Wirtz1924AN....222...21W}.
Redshifts had now become a fundamental issue in
cosmology. In the paper, Wirtz gave his view of the cosmic models of Einstein
and de Sitter; Einstein's model contained a maximum amount of matter, whereas in
de Sitter's model all the mass had been pushed to an unobservable mass horizon,
where the mass was needed to maintain emptiness in the interior. He repeats all
the essential features of de Sitter's model and stresses that in de Sitter's
universe things happen relative to the origin of a coordinate system, but that
every point in the universe can be the origin of that coordinate system. 

Time runs differently, depending on the distance from the origin of
the coordinate system, which is identified with the observer. The slowing down
of time can be seen by the observer as a redshift in the spectral lines. Can
this feature serve to verify de Sitter's theory? Redshifts are known, but the
distances to spiral nebulae are not known at this point. However, if it is
assumed that all spiral nebulae are basically the same, then their apparent
diameters are a measure for their distance. In de Sitter's cosmology, radial
velocities should increase with decreasing apparent diameter. He then looked for
apparent diameters of the objects for which Slipher had given redshifts. He
cites as his sources \cite{Curtis1918PLicO..13....9C} and Pease
(Mt.Wilson Contributions 1919, 1920).\footnote{See
\cite{Pease1920ApJ....51..276P}}
From the text it is clear that
he played around with data in different ways, but the essential result is a list
where he groups the 42 nebulae into 6 groups with $n$ members according
to increasing apparent diameter (Dm= photographic apparent diameter, measured
along the major axis in arc minutes): 

\begin{table}
\begin{tabular}{ccc} 
Log Dm  &    \emph{v} [km s$^{-1}$] &  \emph{n}\\
0.24    &      +827          &   9      \\
0.43    &      +656          &   7      \\
0.66    &      +512          &   8      \\
0.88    &      +555          &   10     \\
1.07    &      +334          &   5      \\
1.71    &       -20          &   3      \\
\end{tabular}
\end{table}

Of course, he was aware that a small apparent diameter may be due to a
smaller than average nebula and not to a large distance. He tried to take that
effect into account and found the logarithmic relationship
$v$ (km s$^{-1}$) $= 2200 - 1200 \cdot log(Dm)$

\emph{``There remains no doubt''}, Wirtz wrote,
``\emph{that the positive radial velocity increases considerably with
increasing distance''}. However, it was later found that his logarithmic
dependence underestimated the gradient. As pointed out by Appenzeller, the
Slipher redshifts were not a statistically representative sample. Only with
difficulty could nebulae with small apparent diameter and large redshifts be
observed \citep{Appenzeller2009}. However, Wirtz was satisfied to have shown the
systematic increase of the nebular redshifts with distance, apparently
confirming de Sitter's world model. There is a lesson for us all here: when
observation matches the predictions of a theoretical model, this does not
constitute proof that the model is correct.

\subsection{Silberstein, Lundmark and Str\"{o}mberg}

In the same year, Ludvik Silberstein and Knut Lundmark also
investigated the relevance of Slipher's data for de Sitter's model
\citep{Silberstein1924MNRAS..84..363S,Lundmark1924MNRAS..84..747L}.\footnote{On
the same subject Silberstein also published several letters
to \emph{Nature}.}
Whereas Wirtz intended to find out whether de Sitter's model
was compatible with observations, Silberstein trusted the model and wanted to
derive a numerical value for $R$, the radius of the universe. From de
Sitter's work he derived his own formula for cosmological redshifts:
\begin{equation}
\frac{d\lambda }{\lambda } =\pm \frac{r}{R}
\end{equation}

Silberstein thus had a formula which also worked for negative velocities. It was
severely criticized by Eddington \citep{Eddington1924Natur.113..746E},
but Silberstein applied the
formula to the study of globular clusters, O-stars and other objects, arriving
at a value of  $6\times10^{12}$ astronomical units for $R$ in 1924.
Today we know that these attempts had to fail.\footnote{
Silberstein's negative sign was also discussed and
criticized by \cite{Lemaitre1925b}.}

\begin{figure}[ht]
\center{\includegraphics[angle=270,scale=0.5]{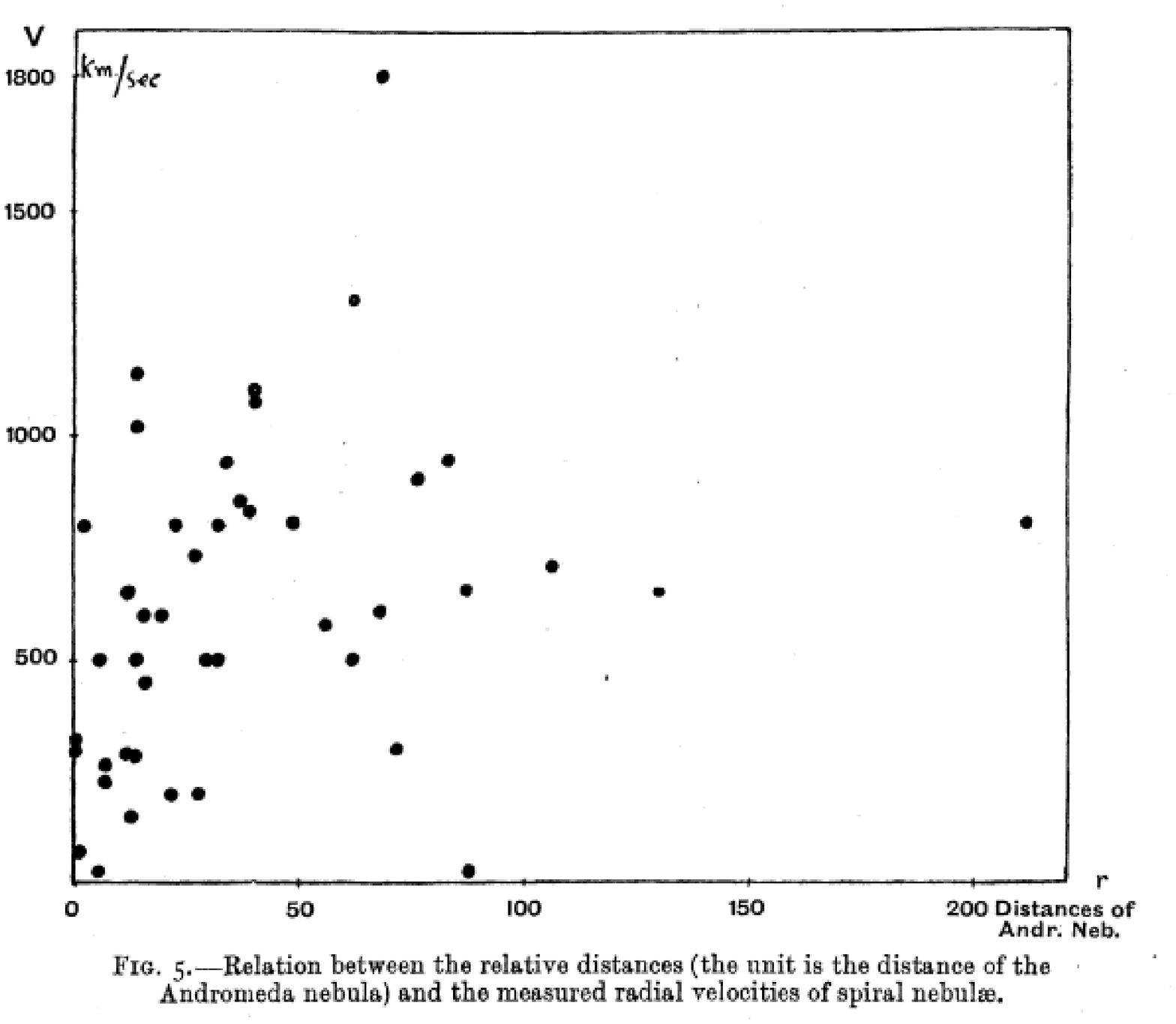}}
\caption{Velocity-distance relation published in
\cite{Lundmark1924MNRAS..84..747L} }\label{nussbaumerfig01}
\end{figure}

Inspired by Silberstein's publications, \cite{Lundmark1924MNRAS..84..747L}
published the
comprehensive study ``The Determination of the Curvature of Space-Time in de
Sitter's World''. In this paper, he stated that his work was based on
``\emph{the wonderful spectrographic work performed at the Lowell Observatory
by Dr. V.M. Slipher.}'' After having severely criticized Silberstein for the
arbitrary choice of his object when deriving $R$, he showed that neither
globular clusters nor stars are much good for determining the curvature of
spacetime, because they are simply too close, he then moved on to spiral
nebulae. Like Wirtz before him, he assumed all spirals to have the same physical
characteristics, such that their apparent angular diameters and magnitudes
depended only on distance. Expressing distance in units of the distance to
Andromeda, he published the diagram shown in
Figure \ref{nussbaumerfig01}. -- This is the first
example of what was later termed a ``Hubble diagram''.

In contrast to Silberstein, Lundmark doesn't believe to have found a
reliable value for $R$ but concludes: ``\emph{we find that there may be
a relation between the two quantities, although not a definite one.''} Actually,
when reading Lundmark one gets the impression that he does not doubt de Sitter's
model, but that he wonders whether the observed nebular motions are not simply
due to normal Doppler shifts. These doubts are understandable considering the
great uncertainties in nebular distances at the time. For example, Ernst \"{O}pik
estimated the distance to Andromeda as 450 kiloparsec
\citep{Opik1922ApJ....55..406O}, while
Lundmark gave a value of 200 kiloparsec, based on the assumption that the
absolute magnitude of Novae at maximum brightness in the mean is the same for
Novae in the Milky Way and in Andromeda (they were not yet aware of supernovae),
and Silberstein quoted even much smaller numbers. These doubts were only
dispelled with Hubble's famous paper read on January 1, 1925 at the American
Astronomical Society meeting in Washington \citep{Hubble1925Obs....48..139H}.
Hubble's distances
were obtained using Cepheid variable stars as standard candles. Although his
result for Andromeda was much less accurate than \"{O}pik's, it was based on a
method that provided the possibility for deriving a consistent set of data for a
large number of nebulae. 

A further study, ``Analysis of radial velocities of globular clusters
and non-galactic nebulae'', came in 1925 from Gustaf Str\"{o}mberg, an observer at
the Mt Wilson Observatory \citep{Stromberg1925ApJ....61..353S}.
The motive for Str\"{o}mberg's study
was twofold; to determine the solar motion and to determine the curvature of
space-time. Str\"{o}mberg stressed the difficulty of determining radial velocities,
but added: ``\emph{\dots  through the perseverance of Professor V.M. Slipher, a
fairly large number of such velocities has been derived.}'' He inserted a table
with radial velocities of globular clusters and non-galactic nebulae, most of
them from Slipher. Note that this table was used by
Lema\^{i}tre in 1927 \footnote{\cite{Lemaitre1927ASSB...47...49L}}
to derive what today is called the ``Hubble constant.''

Comparing observation to theory, Str\"{o}mberg reached a similar
conclusion to Lundmark; ``\emph{In conclusion we may say that we have found no
sufficient reason to believe that there exists any dependence of radial motion
upon distance from the sun.}''

Edwin Hubble was the last to attempt to connect his observations with
de Sitter's theory \citep{Hubble1929PNAS...15..168H}.
Hubble had probably learnt on his visit to
the 1928 IAU General Assembly in Leiden that his ongoing work on the solar
motion might be of relevance to theoreticians. He now did what Wirtz, Lundmark
and Str\"{o}mberg had tried to do before him: to find a relationship between
wavelength shift and distance for the extra-galactic nebulae. Plotting distances
determined by himself against the velocities measured by Slipher, he concluded:
``\emph{The results establish a roughly linear relation between velocities and
distances among nebulae for which velocities have been previously published,
}\dots .''

This was the famous relationship $v=H\cdot r$ , $v=$ velocity
derived from the redshift, $r=$ distance of the nebula, $H=$
factor of proportionality, later called the ``Hubble constant.'' From Slipher's
redshifts and his own distances Hubble
calculated $H=$ 500 (km s$^{-1}$ Mpc$^{-1}$).
Hubble did not know that already two years previously the relationship
$v=H\cdot r$ had been theoretically derived by
\cite{Lemaitre1927ASSB...47...49L}, who
at the same time had also calculated $H$, with practically the same
result as was in 1929 found by Hubble; this will be discussed later.

Hubble's 1929 publication greatly impressed de Sitter. He immediately
realized its importance for advancing the discussion of an appropriate
cosmological model. He analyzed the available observations and discussed them at
a Royal Astronomical Society meeting on 10 January 1930, where Eddington was
present, the discussion is described in \emph{The Observatory}
\citep{deSitter1930Obs....53...37D}.
Between this meeting and the publication of his own findings, which agreed with
Hubble's results, de Sitter learnt about the model of Lema\^{i}tre
\citep{deSitter1930BAN.....5..157D}
and he and Eddington immediately accepted Lema\^{i}tre's
expanding universe.\footnote{The story is told in
detail in \cite{Nussbaumer2009discovering}.}

From 1930 onwards, Hubble and Milton Humason continued Slipher's
pioneering work of redshift observations. They had at their disposition the most
powerful telescope in the world, the Hooker 100-inch on Mount Wilson.

\subsection{Lema\^{i}tre enters the game}

In 1925, Lema\^{i}tre  looked in depth at de Sitter's theoretical
construct and spotted its weak point: de Sitter's solution of the fundamental
equations in the absence of matter introduces a spurious inhomogeneity which is
not simply the mathematical appearance of a center at the origin of coordinates,
but really attributes distinct absolute properties to particular points
\citep{Lemaitre1925a,Lemaitre1925b}.
Although there is great freedom in the choice of the
coordinate system to describe a physical event, this coordinate system must not
by itself change the intrinsic structure of what is described. Yet, de Sitter's
universe is guilty of exactly that misdeed. His line element
$ds= sec(r/R)\cdot dt$ implies that time is running
differently for different values of $r$.
But this violates one of the basic assumptions of cosmology. De
Sitter had chosen a coordinate system that changed the structure of the physical
model. But a coordinate system must not do that; it is there to describe and not
to influence the world. Hence de Sitter's choice of coordinate system needed to
be revised.

In 1925 Lema\^{i}tre introduced a homogeneous division of space and time
and wrote it in the form 

\begin{equation}
ds{}^{2} = R {}^{2} [dt {}^{2} - f(t)\cdot R3],
\end{equation}

\noindent where $R3$ stands for the 3-dimensional Euclidean space. Thus,
for a given time $t$ there is a homogeneous spatial part $R3$,
which, however, with Lema\^{i}tre becomes a function of time. 

In Lema\^{i}tre's coordinate system the radius of space is the same for
any position $r$, but it changes with time $t$. Lema\^{i}tre saw the
implication for cosmology: ``\emph{the radius of space is constant at any
place, but is variable with time,''} and a bit further on: ``\emph{Our
treatment evidences this non-statical character of de Sitter's world which gives
a possible interpretation of the mean receding motion of spiral nebulae}.''  

\begin{figure}[ht]
\center{\plottwo{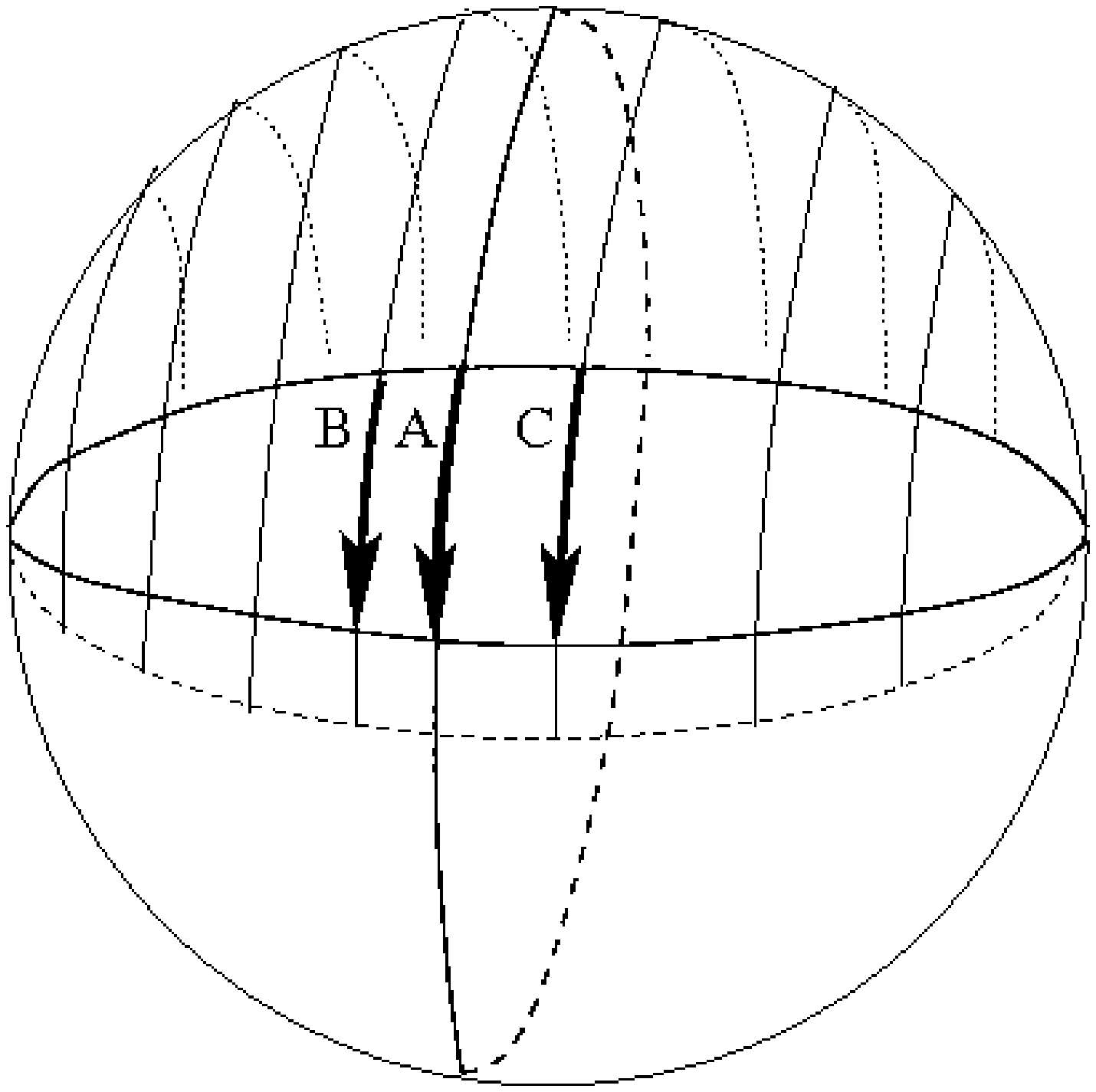}{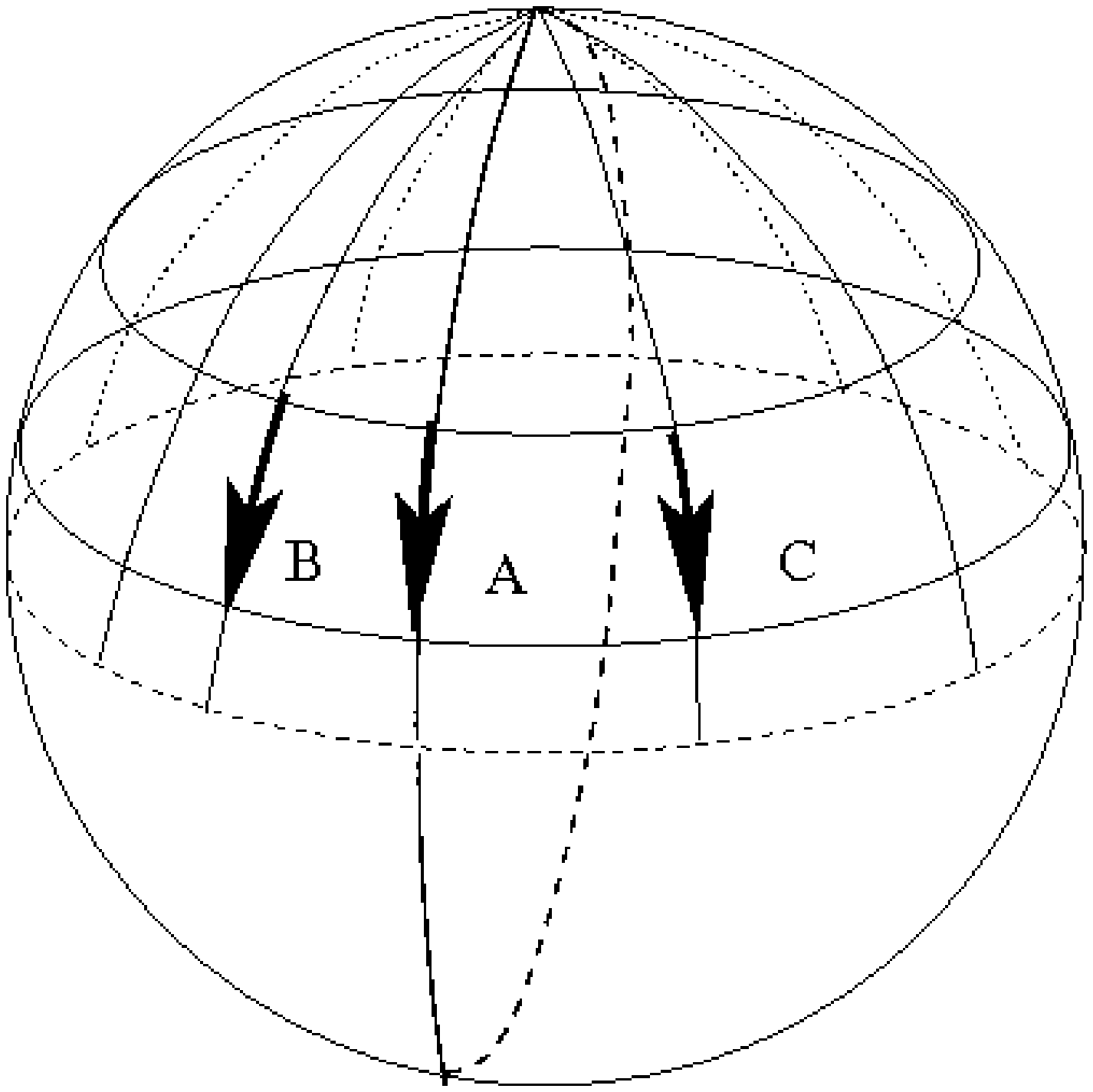}}
\caption{Coordinate systems of de Sitter (left) and Lema\^{i}tre (right).
The two models propose different causes for Slipher's redshifts. For a
detailed description see \cite{Nussbaumer2009discovering}.
}\label{nussbaumerfig02}
\end{figure}

Lema\^{i}tre was not the only one to spot the fallacy in de Sitter's
formalism. In 1922, Kornel Lanczos wrote down a formal solution of a spatially
closed dynamical universe, just as Friedman had done before and Lema\^{i}tre would
do in 1927 \citep{Lanczos1922}. However, unlike
Friedman and Lema\^{i}tre, Lanczos did
not grasp the physical significance and he did not consider a non-stationary
universe. The concept of a static world was deep-rooted. Yet Lanczos' critique
of de Sitter's model was certainly appreciated by Lema\^{i}tre who
refers to it in a footnote in his 1927 paper.

In 1927 Lema\^{i}tre considered the implications of Slipher's redshifts
for theoretical models of the cosmos \citep{Lemaitre1927ASSB...47...49L}.
His key insight
was that the redshifts represent a change in the metric of the universe
between the moment when the light was emitted and when it was observed.
Expressed in more familiar words: redshifts are due to the
expansion of the universe.

In 1927 Lema\^{i}tre writes the line element as
$ds^{2} =-R(t)^{2} d\sigma ^{2} +dt^{2}$,
where $\sigma$ denotes the spatial volume element and $R(t)$ stands for
the radius of curvature of the 3-dimensional space. From this relation, he
derives a relationship between wavelength shifts and distances. For relatively
small distances, $r \ll R$, he obtains $v=H\cdot r$,
where $H$ is positive for an expanding, negative for a contracting, and
zero for a static universe. From Slipher's redshifts, Lema\^{i}tre
concluded that we live in an expanding universe.\footnote{Note that he
did not cite Slipher directly, but obtained the data indirectly from
\cite{Stromberg1925ApJ....61..353S}.}

Noting that the verification of a linear relation between distance and
redshift from observation was not possible from available data due to large
uncertainties in the distances, Lema\^{i}tre concluded that a future verification
was the observer's task. However, he calculated the coefficient $H$, by
taking mean values for $v$ and $r$ from the data of Slipher and
Hubble respectively. Giving equal weight to all observations, he obtained
$H=$575 km s$^{-1}$ Mpc$^{-1}$. Giving less weight to more
distant nebulae resulted in $H=$625 km s$^{-1}$ Mpc$^{-1}$.
As mentioned before, two years later Hubble used
practically the same data, and after having toyed with different data
selections, he opted for $H=$500 km s$^{-1}$ Mpc$^{-1}$
\citep{Hubble1929PNAS...15..168H}.  Thus the
results of the two authors compare favorably with each other. Lema\^{i}tre's
derivation of the numerical value of \emph{H} was omitted in the 1931
translation of his 1927 paper \citep{Lemaitre1931MNRAS..91..483L}.
The omission is due to Lema\^{i}tre
himself \citep[see][]{Livio2011Natur.479..171L}. 

Lema\^{i}tre gave the references for the observational data which entered
his calculation of \emph{H}, however, he did not publish a plot. This was done
much later by Duerbeck and Seitter, as shown in Fig. 3
\citep{Duerbeck2000AcHA...10..120D}.

\begin{figure}[ht]
\center{\includegraphics[angle=270,scale=0.5]{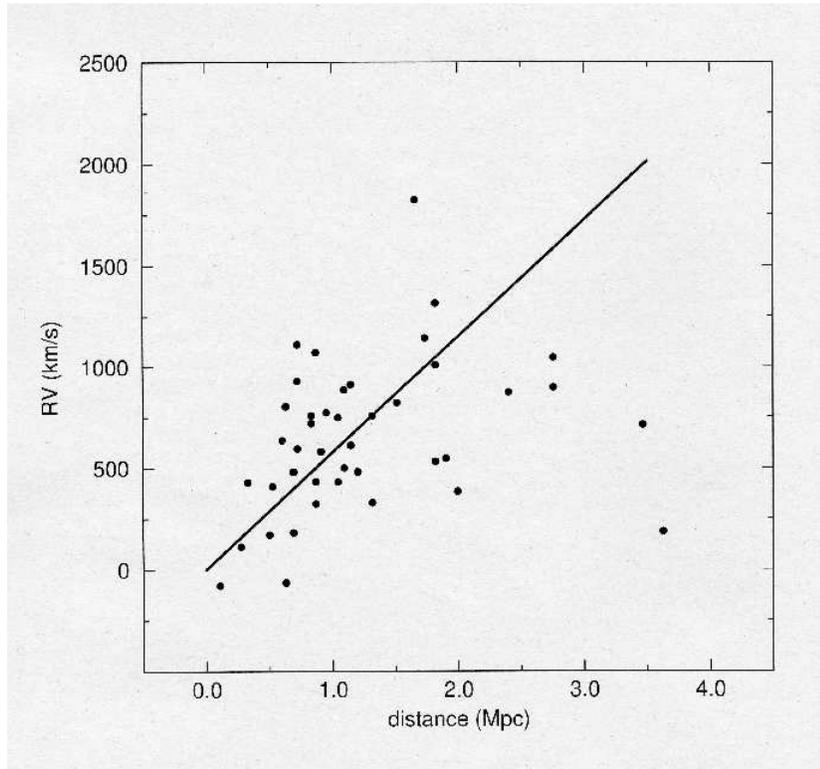}}
\caption{Lema\^{i}tre's velocity-distance relation.
Individual errors in these points are small for velocities but
very large for distances.  Lema\^{i}tre did not publish this
diagram, but used the data points shown. The graph has been
reconstructed by \cite{Duerbeck2000AcHA...10..120D}.}\label{nussbaumerfig03}
\end{figure}

Thus, Slipher's redshifts enter twice in the history of cosmology.
They clearly favored the model of de Sitter over that of Einstein during the
early 1920s, and when Lema\^{i}tre derived a dynamic model in 1927, they
suggested a universe neither static nor shrinking but expanding. 

\section{The dynamic models of Friedman and Lema\^{i}tre}

In 1917, Einstein and de Sitter attempted to describe a static
universe. But in 1922, Friedman showed that Einstein's fundamental equations
also allow dynamic solutions. Einstein took note of Friedman's publication, but
brushed it aside as physically irrelevant. It was only when Lema\^{i}tre, having
spotted de Sitter's violation of the principle of homogeneity of the universe,
found dynamic solutions of Einstein's fundamental equations and combined them
with Slipher's redshifts that the dynamic universe emerged from the mathematical
possibility into physical reality. On whom should we bestow the credit for the
discovery of the expanding universe?

Friedman's publications of 1922 and 1924 showed the author's deep
insight into the cosmological aspects of Einstein's theory of general
relativity. In addition to the static models advocated in 1917 by Einstein and
de Sitter, Friedman gave the mathematical solutions for a dynamic universe,
expanding, contracting or periodic. He also showed that solutions existed not
only for positive but also for negative curvature, and that the universe might
be finite or infinite. Alas, his publications were ignored by the rest of the
scientific community, except for Einstein, who admitted that Friedman's
solutions were mathematically correct (after having refused them initially as
mathematically incorrect), but doubted they were physically significant. Had
Einstein been more receptive, the discovery of an expanding universe could have
occurred many years earlier. 

When we talk about what Friedman did, we also have to mention what he
did not do. Friedman did not suggest in what kind of universe we were living:
was it static as suggested by Einstein and de Sitter, or was it contracting,
forever expanding, or perhaps even oscillating? Friedman did not try to find a
link to existing observations in spite of de Sitter's prediction of wavelength
shifts in his 1917 paper. For this reason Friedman does not qualify as the
discoverer of the expanding universe. This is not to imply that he was not
interested in the practical application of his mathematical findings. He lived
under the very difficult post-revolutionary circumstances of the Soviet Union
which was deliberately isolated by the Western powers. When Friedman published
his first article, access to new astronomical observations had only just begun
to flow again. Thus it may be considered a historical injustice that Western
ignorance of his ground-breaking mathematical work and his untimely death in
September 1925 prevented Friedman from participating actively in seeking the
kind of world we are living in, an undertaking only possible by combining theory
and observations.

When  Lema\^{i}tre restarted his cosmological investigation in 1927, he
had a clear aim. As mentioned above, he had already made the bold step of
adopting the possibility of a dynamic universe in 1925, unaware of the earlier
work of Friedman. In 1927 Lema\^{i}tre was concerned with the enigma of Slipher's
redshifts. His theoretical derivation of the linear velocity-distance
relationship, $v=H\cdot r$, suggested to him that the spectra of the spiral
nebulae held the answer to the question whether we live in a static, contracting
or expanding universe. Thus the title of Lema\^{i}tre's publication of 1927, which
translated from the original French is: \emph{``A homogeneous universe of
constant mass and growing radius, which accounts for the radial velocity of the
extragalactic nebulae''.} Lema\^{i}tre sent a copy to Eddington, who,
however, did not realize at the time that he had in his hands
the solution to a problem which had preoccupied him for many years.
He had also shown it to Einstein, who qualified it as ``physically abominable.''

Eddington and de Sitter, as well as the rest of the astronomical
community only took note of Lema\^{i}tre's work at the beginning of 1930 after
having heard about Hubble's observational finding of a linear velocity-distance
relationship, but they immediately welcomed it as the solution to the
long-standing cosmological problem of Slipher's redshifts and de Sitter's
incomprehensible model. As mentioned before, if the theoretical physicists and
the astronomical community had realized the potential of Friedman's work, it all
might have happened before, but it didn't. And history talks about what happened
and not what might have happened; thus credit for the discovery of the expanding
universe goes to Lema\^{i}tre.

How did the expansion start? When Eddington showed that Einstein's
static universe was unstable, he suggested that such a pseudo-static universe
might have been the original status of the universe
\citep{Eddington1930MNRAS..90..668E}. This
opinion was shared by Lema\^{i}tre. However, both met difficulties when trying to
explain how such an equilibrium slid into expansion.\footnote{There was a
lively debate on this point in the \emph{Monthly Notices} of 1930 and 1931, see
\citet[][p.165]{Nussbaumer2009discovering}.}

Inspired by the process of radioactivity, Lema\^{i}tre in 1931 replaced the
mathematical singularity at time zero by a primeval atom containing all the
matter of the universe. This highly unstable atom would then decay by a process
of super-radioactivity. This suggestion, published in \emph{Nature, }marks the
precursor of today's big bang model \citep{Lemaitre1931Natur.127..706L}.
On philosophical grounds
Eddington was not happy with the idea, but they both agreed that the
cosmological constant, $\Lambda $, was a fundamental force in nature, and that
the history of expansion was primarily determined by the strength of $\Lambda $
relative to the gravitational force. Lema\^{i}tre's equations were well tailored
to deal with that process. The work of Friedman did not enter this cosmological
discussion, as the motive for introducing the big bang was not mathematical but
physical. Thus, Lema\^{i}tre did not need to borrow anything from Friedman that
was not already contained in his own 1925 and 1927 papers. In addition, in
November 1933 Lema\^{i}tre presented a contribution ``Evolution of the expanding
universe'' to the National Academy of Sciences, where he associated $\Lambda$
with vacuum energy, in agreement with today's interpretation
\citep{Lemaitre1934PNAS...20...12L}.

In 1932, Einstein and de Sitter published a model of the expanding
universe that did not contain a cosmological constant
\citep{Einstein1932PNAS...18..213E}; this became the standard
model for many years. They wrote: \emph{``Dr.  Heckmann has
pointed out that the non-static solutions of the field equations of
the general theory of relativity with constant density do not necessarily imply
a positive curvature of three-dimensional space, but that this curvature may
also be negative or zero.''} Heckmann's publication did not refer to Friedman.
However, it discussed solutions which are already implicit in Friedman's work.
Whether Heckmann profited from Friedman without referring to him, we do not
know. Thus, it may be that Friedman had a direct influence on the Einstein-de
Sitter model. However, this model also follows from Lema\^{i}tre's solution,
if one choses $\Lambda $=0, and if the particle density corresponds to the
critical density, which was assumed by Einstein and de Sitter. As we now know,
this assumption is definitely not fulfilled.

Einstein thought highly of Friedman. He implicitly tells us his reason
in his 1931 paper, where he converted to the expanding universe (translated from
German): ``\emph{Several investigators have tried to cope with the new facts
by using a spherical space whose radius, P, is variable in time. The first who,
uninfluenced by observations, tried this way was A. Friedman}''
\citep{Einstein1931SPAW.......235}.
Einstein admired Friedman for having found dynamical solutions without
being directed by observations. He could probably have kicked himself for not
having spotted these solutions himself, and he may have felt guilty for having
pushed aside Friedman in such a high-handed way.

The contrast between Friedman and Lema\^{i}tre may be seen in the titles
of their main publications. Friedman's titles were ``On the curvature of Space''
\citep{Friedman1922ZPhy...10..377F} and ``About the possibility of a world
of constant negative curvature of space'' \citep{Friedman1924ZPhy...21..326F},
whereas \cite{Lemaitre1927ASSB...47...49L} published about
``A homogeneous universe of constant mass and growing radius
accounting for the radial velocities of the
extra-galactic nebulae.'' The two scientists obviously had quite different
priorities. It makes little sense to blame either of them for not having added
to his work what the other had done.

Friedman's work showed a fundamental insight into the cosmological
content of Einstein's fundamental equations. It is recognized that he gave all
the cosmologically relevant mathematical solutions of Einstein's fundamental
equations, including the possibility of a dynamical expanding, shrinking or
periodic universe. Once Lema\^{i}tre became aware of Friedman's work late in 1927,
he always acknowledged that Friedman was the first to find the mathematical
solution of an expanding universe \citep[e.g.][]{Lemaitre1931MNRAS..91..483L}.
Friedman's work has
also been fully acknowledged by the scientific community; however, it would be a
historical distortion to build him up as the discoverer of the expanding
universe. Friedman never debated why, of all his mathematical solutions, the
expanding universe should be the one in which we live; this was done by
Lema\^{i}tre. A further discussion of these points can be found in the book
\emph{Discovering the expanding universe} \citep{Nussbaumer2009discovering}.

\acknowledgments I am grateful to Michael Way and an anonymous referee for
very constructive suggestions.

\bibliography{nussbaumer}

\begin{thebibliography}{}
\expandafter\ifx\csname natexlab\endcsname\relax\def\natexlab#1{#1}\fi
\expandafter\ifx\csname url\endcsname\relax
  \def\url#1{\texttt{#1}}\fi
\expandafter\ifx\csname urlprefix\endcsname\relax\def\urlprefix{URL }\fi
\providecommand{\eprint}[2][]{\url{#2}}

\bibitem[{Appenzeller(2009)}]{Appenzeller2009}
Appenzeller, I. 2009, {Carl Wirtz und die Hubble-Beziehung}, Sterne und
  Weltraum, 11, 44

\bibitem[{{Curtis}(1918)}]{Curtis1918PLicO..13....9C}
{Curtis}, H.~D. 1918, {Descriptions of 762 Nebulae and Clusters Photographed
  with the Crossley Reflector}, Publications of Lick Observatory, 13, 9

\bibitem[{{de Sitter}(1917)}]{deSitter1917MNRAS..78....3D}
{de Sitter}, W. 1917, {Einstein's Theory of Gravitation and its Astronomical
  Consequences. Third paper}, \mnras, 78, 3

\bibitem[{{de Sitter}(1930{\natexlab{a}})}]{deSitter1930Obs....53...37D}
--- 1930{\natexlab{a}}, {Notes}, The Observatory, 53, 37

\bibitem[{{de Sitter}(1930{\natexlab{b}})}]{deSitter1930BAN.....5..157D}
--- 1930{\natexlab{b}}, {On the Magnitudes, Diameters and Distances of the
  Extragalactic Nebulae and their Apparent Radial Velocities (Errata: 5 V,
  230)}, {Bulletin of the Astronomical Institutes of the Netherlands}, 5, 157

\bibitem[{{Duerbeck} \& {Seitter}(2000)}]{Duerbeck2000AcHA...10..120D}
{Duerbeck}, H.~W., \& {Seitter}, W.~C. 2000, {In Edwin Hubble's Shadow: Early
  Investigations on the Expansion of the Universe}, Acta Historica Astronomiae,
  10, 120

\bibitem[{{Eddington}(1917)}]{Eddington1917MNRAS..77..375}
{Eddington}, A.~S. 1917, {Society Business: Progress and Present State of the
  Society; Report of the Council to the Ninety-Seventh Annual General Meeting
  of the Society}, \mnras, 77, 375

\bibitem[{{Eddington}(1923)}]{Eddington1923mtr..book.....E}
--- 1923, {The Mathematical Theory of Relativity} (Cambridge University Press)

\bibitem[{{Eddington}(1924)}]{Eddington1924Natur.113..746E}
--- 1924, {Radial Velocities and the Curvature of Space-Time}, \nat, 113, 746

\bibitem[{{Eddington}(1930)}]{Eddington1930MNRAS..90..668E}
--- 1930, {On the Instability of Einstein's Spherical World}, \mnras, 90, 668

\bibitem[{{Einstein}(1917)}]{Eddington1917SPAW.......142E}
{Einstein}, A. 1917, {Kosmologische Betrachtungen zur Allgemeinen
  Relativit{\"a}tstheorie}, Sitzungsberichte der K{\"o}niglich Preu{\ss}ischen
  Akademie der Wissenschaften (Berlin), Seite 142-152

\bibitem[{{Einstein}(1931)}]{Einstein1931SPAW.......235}
--- 1931, {Zum Kosmologischen Problem der Allgemeinen Relativitätstheorie},
  Sitzungsberichte der K{\"o}niglich Preu{\ss}ischen Akademie der
  Wissenschaften (Berlin), Seite 235-237

\bibitem[{{Einstein} \& {de Sitter}(1932)}]{Einstein1932PNAS...18..213E}
{Einstein}, A., \& {de Sitter}, W. 1932, {On the Relation between the Expansion
  and the Mean Density of the Universe}, Proceedings of the National Academy of
  Science, 18, 213

\bibitem[{{Friedman}(1922)}]{Friedman1922ZPhy...10..377F}
{Friedman}, A. 1922, {{\"U}ber die Kr{\"u}mmung des Raumes}, Zeitschrift
  f\"{u}r Physik, 10, 377

\bibitem[{{Friedman}(1924)}]{Friedman1924ZPhy...21..326F}
--- 1924, {{\"U}ber die M{\"o}glichkeit einer Welt mit Konstanter Negativer
  Kr{\"u}mmung des Raumes}, Zeitschrift f\"{u}r Physik, 21, 326

\bibitem[{{Hubble}(1929)}]{Hubble1929PNAS...15..168H}
{Hubble}, E. 1929, {A Relation between Distance and Radial Velocity among
  Extra-Galactic Nebulae}, Proceedings of the National Academy of Science, 15,
  168

\bibitem[{{Hubble}(1925)}]{Hubble1925Obs....48..139H}
{Hubble}, E.~P. 1925, {Cepheids in Spiral Nebulae}, The Observatory, 48, 139

\bibitem[{{Lanczos}(1922)}]{Lanczos1922}
{Lanczos}, K. 1922, {Bemerkungen zur de Sitterschen Welt}, Physikalische
  Zeitschrift, 23, 539

\bibitem[{{Lema\^{i}tre}(1925{\natexlab{a}})}]{Lemaitre1925b}
{Lema\^{i}tre}, G. 1925{\natexlab{a}}, {Note on de Sitter’s Universe},
  Physical Review, 25 Ser.II, 903

\bibitem[{{Lema\^{i}tre}(1925{\natexlab{b}})}]{Lemaitre1925a}
--- 1925{\natexlab{b}}, {Note on de Sitter’s Universe}, Journal of
  Mathematics and Physics, 4, 188

\bibitem[{{Lema{\^i}tre}(1927)}]{Lemaitre1927ASSB...47...49L}
{Lema{\^i}tre}, G. 1927, {Un Univers Homog{\`e}ne de Masse Constante et de
  Rayon Croissant Rendant Compte de la Vitesse Radiale des N{\'e}buleuses
  Extra-Galactiques}, Annales de la Societe Scietifique de Bruxelles, 47, 49

\bibitem[{{Lema{\^i}tre}(1931{\natexlab{a}})}]{Lemaitre1931MNRAS..91..483L}
--- 1931{\natexlab{a}}, {A Homogeneous Universe of Constant Mass and Increasing
  Radius accounting for the Radial Velocity of Extra-galactic Nebulae}, Monthly
  Notices of the Royal Astronomical Society, 91, 483

\bibitem[{{Lema{\^i}tre}(1931{\natexlab{b}})}]{Lemaitre1931Natur.127..706L}
--- 1931{\natexlab{b}}, {The Beginning of the World from the Point of View of
  Quantum Theory}, \nat, 127, 706

\bibitem[{{Lema\^{i}tre}(1934)}]{Lemaitre1934PNAS...20...12L}
{Lema\^{i}tre}, G. 1934, {Evolution of the Expanding Universe}, Proceedings of
  the National Academy of Science, 20, 12

\bibitem[{{Livio}(2011)}]{Livio2011Natur.479..171L}
{Livio}, M. 2011, {Lost in Translation: Mystery of the Missing Text Solved},
  \nat, 479, 171

\bibitem[{{Lundmark}(1924)}]{Lundmark1924MNRAS..84..747L}
{Lundmark}, K. 1924, {The Determination of the Curvature of Space-Time in de
  Sitter's World}, \mnras, 84, 747

\bibitem[{{Nussbaumer} \& {Bieri}(2009)}]{Nussbaumer2009discovering}
{Nussbaumer}, H., \& {Bieri}, L. 2009, Discovering the Expanding Universe
  (Cambridge University Press).
  \urlprefix\url{http://books.google.ch/books?id=RaNOJkQ4l14C}

\bibitem[{{\"{O}pik}(1922)}]{Opik1922ApJ....55..406O}
{\"{O}pik}, E. 1922, {An Estimate of the Distance of the Andromeda Nebula},
  \apj, 55, 406

\bibitem[{{Pease}(1920)}]{Pease1920ApJ....51..276P}
{Pease}, F.~G. 1920, {Photographs of Nebulae with the 60-inch Reflector,
  1917-1919}, \apj, 51, 276

\bibitem[{{Silberstein}(1924)}]{Silberstein1924MNRAS..84..363S}
{Silberstein}, L. 1924, {The Curvature of de Sitter's Space-Time Derived from
  Globular Clusters}, \mnras, 84, 363

\bibitem[{{Slipher}(1912)}]{Slipher1912LowOB...2...26S}
{Slipher}, V.~M. 1912, {On the Spectrum of the Nebula in the Pleiades}, Lowell
  Observatory Bulletin, 2, 26

\bibitem[{{Slipher}(1913)}]{Slipher1913LowOB...2...56S}
--- 1913, {The Radial Velocity of the Andromeda Nebula}, Lowell Observatory
  Bulletin, 2, 56

\bibitem[{{Slipher}(1915)}]{Slipher1915PA.....23...21S}
--- 1915, {Spectrographic Observations of Nebulae}, Popular Astronomy, 23, 21

\bibitem[{{Slipher}(1917)}]{Slipher1917PAPhS..56..403S}
--- 1917, {Nebulae}, Proceedings of the American Philosophical Society, 56, 403

\bibitem[{{Str\"{o}mberg}(1925)}]{Stromberg1925ApJ....61..353S}
{Str\"{o}mberg}, G. 1925, {Analysis of Radial Velocities of Globular Clusters
  and Non-Galactic Nebulae}, \apj, 61, 353

\bibitem[{{Wirtz}(1922)}]{Wirtz1922AN....215..349W}
{Wirtz}, C. 1922, {Einiges zur Statistik der Radialbewegungen von Spiralnebeln
  und Kugelsternhaufen}, Astronomische Nachrichten, 215, 349

\bibitem[{{Wirtz}(1924)}]{Wirtz1924AN....222...21W}
--- 1924, {De Sitters Kosmologie und die Radialbewegungen der Spiralnebel},
  Astronomische Nachrichten, 222, 21

\end{thebibliography}

\end{document}